\documentclass[]{jfm}
\usepackage{graphicx}
\usepackage{newtxtext}
\usepackage{newtxmath}
\usepackage{natbib}
\usepackage{hyperref}
\usepackage[dvipsnames]{xcolor} 
\hypersetup{
    colorlinks = true,
    urlcolor   = blue,
    citecolor  = black,
}

\newcommand{\RomanNumeralCaps}[1]
\linenumbers

\title{Energy pathways in large- and small-scale convection-driven dynamos.}

\author{Souvik Naskar\aff{1}, 
  Anikesh Pal\aff{1}
  \corresp{\email{pala@iitk.ac.in}}}

\affiliation{\aff{1}Department of Mechanical Engineering, Indian Institute of Technology, Kanpur 208016, India}

\begin{document}
\maketitle

\begin{abstract}
We investigate the energy pathways between the velocity and the magnetic fields in a rotating plane layer dynamo driven by Rayleigh-B\'enard convection using direct numerical simulations. The kinetic and magnetic energies are divided into mean and turbulent components to study the production, transport, and dissipation associated with large and small-scale dynamos. This energy balance-based characterization reveals distinct mechanisms for large- and small-scale magnetic field generation in dynamos, depending on the nature of the velocity field and the conditions imposed at the boundaries. 
\end{abstract}



\section{Introduction}\label{sec:intro}

Hydromagnetic dynamo action is the commonly accepted source of magnetic fields in planets and stars. We explore the energy balance in a plane layer convection-driven Childress-Soward (CS) dynamo \citep{childress_1972}. In this simple Cartesian dynamo model, driven by rotating Rayleigh-B\'enard convection (RBC), the velocity field provides energy to the magnetic field to sustain the dynamo action against Joule dissipation. Our analysis distinguishes between the part of the kinetic energy that produces the horizontally-averaged large-scale mean magnetic field and the small-scale turbulent magnetic field. Based on this decomposition, we derive the kinetic and magnetic energy budgets and characterize a dynamo with both large- and small-scale magnetic fields in terms of their energy conversion mechanisms. \\

Dynamos are generally classified based on the scale of the magnetic field generated by them \cite{brandenburg_2005}. The magnetic field for a large-scale dynamo has a length scale larger than the velocity field, whereas the length scale of the magnetic field is smaller than the velocity field for a small-scale dynamo. For a plane layer dynamo, this classification is apparent from the fraction of energy in the mean magnetic field to the total magnetic energy \citep{tilgner_2012}. The magnetic Reynolds number at the convective scale $\widetilde{Rm}$, representing the dominance of electromagnetic induction over ohmic diffusion, is one of the important diagnostic quantities that decides this mean energy fraction. Large-scale dynamos with high mean energy fraction, $O(0.1)$ has been found to operate at low magnetic Reynolds numbers $\widetilde{Rm}\lesssim13$ \citep{tilgner_2012}, where kinetic helicity has been proposed as a driving mechanism. Conversely, higher values of $\widetilde{Rm}$ lead to small-scale magnetic field generation driven by the stretching of magnetic field lines by the velocity field, with comparatively lower mean energy fraction. However, a recent study \cite{yan_2022b} demonstrates the existence of large-scale dynamos, despite the presence of a strongly turbulent velocity field with low helicity. In the present investigation, we use the kinetic and magnetic energy budgets to understand the various mechanisms associated with large-and small-scale magnetic field generation.\\

\section{Method}\label{sec:method}

\subsection{Governing Equations}\label{sec:govequ}
In this study, the CS dynamo is driven by a classical Rayleigh-B\'enard convection setup with a plane layer of incompressible, electrically conducting, Boussinesq fluid kept between two parallel plates with a distance $d$ and temperature difference $\Delta T$, where the lower plate is hotter than the upper plate. The system rotates with a constant angular velocity $\Omega$ about the vertical axis, anti-parallel to the gravity $g$. This fluid has the kinematic viscosity $\nu$, thermal diffusivity $\kappa$, adiabatic volume expansion coefficient $\alpha$, and magnetic diffusivity $\eta$. The Navier-Stokes equations, coupled with the energy equation, the Maxwell equation, and the solenoidal field conditions, govern the velocity, pressure, temperature, and the magnetic field $\{u_i, p, \theta, B_i\}$ \citep{naskar_2022a,naskar_2022b} as presented below. 


\begin{equation}\label{eqn:solenoidal_nd}
    \frac{\partial u_j}{\partial x_j}=
    \frac{\partial B_j}{\partial x_j}=0,
\end{equation}
\begin{equation}\label{eqn:momentum_nd}
\begin{split}
    \frac{\partial u_i}{\partial t}+
    u_j\frac{\partial u_i}{\partial x_j}=
    -\frac{\partial p}{\partial x_i}+
    \frac{1}{E}\sqrt{\frac{Pr}{Ra}}\epsilon_{ij3} u_j\hat{e_3}+
    B_j\frac{\partial B_i}{\partial x_j}+
    \theta\delta_{i3}+
    \sqrt{\frac{Pr}{Ra}}\frac{\partial^{2} u_{i}}{\partial x_j\partial x_j},
\end{split}
\end{equation}
\begin{equation}\label{eqn:energy_nd}
    \frac{\partial \theta}{\partial t}+
    u_j\frac{\partial \theta}{\partial x_j}=
    \frac{1}{\sqrt{RaPr}}\frac{\partial^2\theta}{\partial x_j\partial x_j},
\end{equation}
\begin{equation}\label{eqn:maxwell_nd}
    \frac{\partial B_i}{\partial t}+
    u_j\frac{\partial B_i}{\partial x_j}=
    B_j\frac{\partial u_i}{\partial x_j}+
    \sqrt{\frac{Pr}{Ra}}\frac{1}{Pr_m}\frac{\partial^{2} B_{i}}{\partial x_j\partial x_j}.
\end{equation}

The governing non-dimensional parameters are the Rayleigh number ($Ra=g\alpha\Delta T d^3/\kappa\nu$) and Ekman number ($E=\nu/2\Omega d^2$) representing the thermal forcing and rotation rates, respectively, whereas, the thermal and magnetic Prandtl numbers ($Pr=\nu/\kappa$ and $Pm=\nu/\eta$) are the properties of the fluid. The current setup is a local approximation to the astrophysically more relevant spherical shell dynamo models \citep{tilgner_2012}. In the horizontal directions ($x_{1},x_{2}$) periodic boundary conditions are applied. Both no-slip and free-slip boundary conditions have been used in the vertical direction ($x_3$).

\begin{equation}\label{eqn:ubc}
\begin{split}
    u_{1}=u_{2}=u_{3}=0 \; \textrm{ at } \; x_{3}=\pm1/2 \qquad\text{(no-slip)} \\
   \frac{\partial u_{1}}{\partial x_{3}}=\frac{\partial  u_{2}}{\partial x_{3}}=0,\  u_{3}=0 \  \textrm{ at } \  x_{3}=\pm1/2 \qquad\text{(free-slip)}.    
\end{split}
\end{equation} 

Thermal boundary conditions are isothermal with unstable temperature gradients.

\begin{equation}\label{eqn:tbc}
\begin{split}
    \quad \theta=1/2 \; \textrm{ at } \; x_{3}=-1/2, \quad \theta=-1/2 \; \textrm{ at } \; x_{3}=1/2
\end{split}
\end{equation}

Perfectly conducting and pseudo-vacuum boundary conditions have been used for the magnetic field.

\begin{equation}\label{eqn:mbc}
\begin{split}
   \frac{\partial B_{1}}{\partial x_{3}}=\frac{\partial  B_{2}}{\partial x_{3}}=B_{3}=0 \  \textrm{ at } \  x_{3}=\pm1/2 \qquad\text{(perfectly conducting)} \\
   B_{1}=B_{2}=\  \frac{\partial B_{3}}{\partial x_{3}}=0 \  \textrm{ at } \  x_{3}=\pm1/2 \qquad\text{(pseudo-vacuum)}   
\end{split}
\end{equation} 

\subsection{Energy Budget}\label{sec:budget}

We perform a Reynolds decomposition of the variables into mean and fluctuating parts such that $\phi(x,y,z,t)=\Bar{\phi}(z,t)+\phi '(x,y,z,t)$ where $\phi=\{u_i, p, \theta, B_i\}$. Here, the over-bar denotes an average over the horizontal directions\citep{tilgner_2012,naskar_2022b}. The kinetic energy($K=1/2u_iu_i$) and magnetic energy ($M=1/2B_iB_i$) are also divided into the mean ($\mathfrak{K}$ and $\mathfrak{M}$) and turbulent ($\mathcal{K}$ and $\mathcal{M}$) components, and are presented in equations \ref{eqn:TKE_budget}-\ref{eqn:MME_budget}. These equations can be derived in three steps: (1) horizontally averaging equations \ref{eqn:momentum_nd} and \ref{eqn:maxwell_nd} to get the mean momentum and mean Maxwell equations respectively, and then subtracting these mean equations from \ref{eqn:momentum_nd} and \ref{eqn:maxwell_nd} to get the fluctuating parts of momentum and Maxwell equations (2) multiplying the mean velocity and mean magnetic field with mean momentum and mean Maxwell equations respectively, to get evolution equations for $\mathfrak{K}$ and $\mathfrak{M}$, (3) multiplying the fluctuating velocity and magnetic fields with fluctuating parts of momentum and Maxwell equations respectively, and averaging them to get evolution equations for $\mathcal{K}$ and $\mathcal{M}$. This decomposition into mean and turbulent parts of the energies is the primary distinctive feature of the present study from an earlier budget analysis of a dynamo \citep{brandenburg_1996}.
The turbulent kinetic energy (TKE) \citep{naskar_2022b} evolves as: 
\begin{equation}\label{eqn:TKE_budget}
    d\mathcal{K}/dt=\mathcal{S}+\mathcal{B}-\mathcal{D}-\partial\mathcal{T}_{j}/\partial x_j+\mathcal{P},
\end{equation}
where $\mathcal{K}=1/2\overline{u^{\prime}_i{}u^{\prime}_i{}}$ is the TKE, $\mathcal{S}=-\overline{u^{\prime}_iu^{\prime}_j}\partial\Bar{u_i}/\partial x_j$ is the production of TKE by mean shear, $\mathcal{B}=\overline{u^{\prime}_3\theta^{\prime}}$ is the conversion of available potential energy (APE) to TKE by the turbulent buoyancy flux \citep{gayen_2013}, $\mathcal{D}=\sqrt{Pr/Ra}\ \overline{\partial u^{\prime}_i/\partial x_j\partial u^{\prime}_i/\partial x_j}$ is the viscous dissipation which converts TKE to internal energy (IE), $\partial\mathcal{T}_{j}/\partial x_j$ is the redistribution of TKE \citep{petschel_2015}, representing the divergence of the TKE flux $\mathcal{T}_j=\mathcal{T}_p+\mathcal{T}_t+\mathcal{T}_v+\mathcal{T}_M$. The components of the TKE flux are the pressure flux, $\mathcal{T}_p=\overline{u^{\prime}_jp^{\prime}}$, the turbulent flux $\mathcal{T}_t=\frac{1}{2}\overline{u^{\prime}_iu^{\prime}_iu^{\prime}_j}$, the viscous flux $\mathcal{T}_v=-\sqrt{Pr/Ra}\partial \mathcal{K}/\partial x_j$, and the magnetic flux $\mathcal{T}_M=-\overline{B_j}\ \overline{u^{\prime}_i{}B^{\prime}_i}-\overline{u^{\prime}_iB^{\prime}_iB^{\prime}_j}$. 

In equation \ref{eqn:TKE_budget}, the term $\mathcal{P}$ represents the production of $\mathcal{K}$ due to work done by the Lorentz force on the flow field. It can be further divided into three components $\mathcal{P}=-\mathcal{P}_{1}+\mathcal{P}_{2}-\mathcal{P}_{3}$, where $\mathcal{P}_{1}=\overline{B}_j\overline{B^{\prime}_i\partial u^{\prime}_i/\partial x_j}$ is the production of TME from TKE due to work done by the fluctuating strain rate on the mean magnetic field, $\mathcal{P}_{2}=\overline{u^{\prime}_iB^{\prime}_j}\partial \overline{B}_i/\partial x_j$ signifies the production of TKE due to mean magnetic field gradient and $\mathcal{P}_{3}=\overline{B^{\prime}_iB^{\prime}_j\partial u^{\prime}_i/\partial x_j}$ represents the amplification (or attenuation) of the magnetic energy, due to work done by stretching (or squeezing) of small-scale magnetic field lines by the fluctuating velocity gradients.

The mean kinetic energy (MKE) budget is expressed as,
\begin{equation}\label{eqn:MKE_budget}
    d\mathfrak{K}/dt=-\mathcal{S}+\mathfrak{B}-\mathfrak{D}-\partial\mathfrak{T}_{j}/\partial x_j+\mathfrak{P} 
\end{equation}
where $\mathfrak{K}=1/2\overline{u}_i\overline{u}_i$ is the MKE, $\mathfrak{B}=\overline{u}_3\overline{\theta}$ is the mean buoyancy flux, $\mathfrak{D}=\sqrt{Pr/Ra}\partial \overline{u}_i/\partial x_j\partial \overline{u}_i\partial x_j$ is the mean viscous dissipation, and $\partial\mathfrak{T}_{j}/\partial x_j$ is the divergence of the MKE flux  $\mathfrak{T}_j=\overline{u_j}\overline{p}+1/2\overline{u^{\prime}_iu^{\prime}_j}{\overline{u}_i}-\sqrt{Pr/Ra}\partial \mathfrak{K}/\partial x_j-\overline{B^{\prime}_iB^{\prime}_j}{\overline{u}_i}-\overline{B}_i\ \overline{B}_j\overline{u}_i$. Here $\mathfrak{P}=-\mathcal{P}_{4}-\mathcal{P}_{5}$, where,
$\mathcal{P}_{4}=\overline{B^{\prime}_iB^{\prime}_j}\partial \overline{u}_i/\partial x_j$ represents the work done by the mean shear on the fluctuating component of the magnetic fields whereas, the work done on the mean component of the magnetic fields is represented by $\mathcal{P}_{5}=\overline{B}_i\overline{B}_j\partial \overline{u}_i/\partial x_j$.


The equations for the evolution of the turbulent and the mean magnetic energies are presented below. As the Maxwell equation \citep{naskar_2022a,naskar_2022b} is similar to the vorticity transport equation, the magnetic energy equations resemble enstrophy transport equations\citep{lumley_1997}. 
The turbulent magnetic energy (TME) budget equation is given below.
\begin{equation}\label{eqn:TME_budget}
    d\mathcal{M}/dt=-\mathcal{D}^{M}-\partial\mathcal{T}^{M}_{j}/\partial x_j+\mathcal{P}^{M}
\end{equation}
where, $\mathcal{M}=1/2\overline{B^{\prime}_i{}B^{\prime}_i{}}$ is the TME, $\mathcal{D}^{M}=1/Pm\sqrt{Pr/Ra}\ \overline{\partial B^{\prime}_i/\partial x_j\partial B^{\prime}_i/\partial x_j}$ is the Joule dissipation, $\partial\mathcal{T}^{M}_{j}/\partial x_j$ is the redistribution of TME flux given by $\mathcal{T}^{M}_j=1/2\overline{B^{\prime}_iB^{\prime}_iu^{\prime}_j}-1/Pm\sqrt{Pr/Ra}\partial \mathcal{M}/\partial x_j$. Here, the energy exchange terms can be expressed as $\mathcal{P}^{M}=\mathcal{P}_{1}+\mathcal{P}_{3}+\mathcal{P}_{4}-\mathcal{P}_{6}$, where $\mathcal{P}_{6}=\overline{u^{\prime}_jB^{\prime}_i}\partial \overline{B}_i/\partial x_j$ exchange energy between the turbulent and mean magnetic fields. We can anticipate its appearance in the mean magnetic energy (MME) equations with an opposite sign.
\begin{equation}\label{eqn:MME_budget}
    d\mathfrak{M}/dt=-\mathfrak{D}^{M}-\partial\mathfrak{T}^{M}_{j}/\partial x_j+\mathfrak{P}^{M}
\end{equation}
where, $\mathfrak{M}=1/2\overline{B}_i\overline{B}_i$ is the MME, $\mathfrak{D}^{M}=1/Pm\sqrt{Pr/Ra}\partial \overline{B}_i/\partial x_j\partial \overline{B}_i/\partial x_j$ is the mean Joule dissipation, and $\partial\mathfrak{T}^{M}_{j}\partial x_j$ is the divergence of MME flux $\mathfrak{T}^{M}_j=\overline{B_i}\ \overline{u^{\prime}_j{}B^{\prime}_i}-\overline{B_i}\ \overline{u^{\prime}_i{}B^{\prime}_j}-1/Pm\sqrt{Pr/Ra}\frac{\partial \mathfrak{M}}{\partial x_j}$. Energy exchange terms in the MME equation can be expressed as $\mathfrak{P}^{M}=-\mathcal{P}_{2}+\mathcal{P}_{5}+\mathcal{P}_{6}$.

\begin{figure}
\centering
\includegraphics[width=0.8\textwidth,trim={4cm 2cm 2cm 2cm},clip]{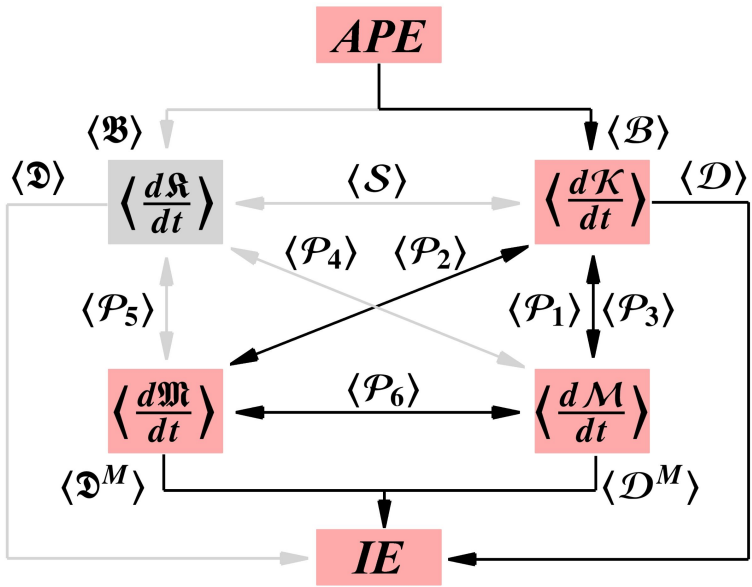}
\caption{Energy pathways between the kinetic and magnetic energies in a convection-driven dynamo. The energetic terms shown here are volume-averaged as indicated by angular brackets $\langle.\rangle$. The energy pathways marked in grey are negligible in the absence of a mean flow.}
\label{fig:energyflow}
\end{figure}

We plot an energy pathway diagram by averaging the terms in the energy budget equations in the vertical direction to obtain the volume averaged quantities, as shown in figure \ref{fig:energyflow}. The transport terms become negligible due to this volume-averaging, and we can demonstrate the conversion paths between KE and ME sustaining the dynamo action. The APE is converted to MKE and TKE via the mean and turbulent part of the buoyancy flux \citep{gayen_2013} respectively. The shear production term, $\mathcal{S}$, may also produce TKE in the presence of a mean shear. The small-scale turbulent flow can exchange energy with the small-scale magnetic field through $\mathcal{P}_1$ or $\mathcal{P}_3$ or both, or with the large-scale mean magnetic field through $\mathcal{P}_2$. The large-scale flow can exchange energy with the small-scale and large-scale magnetic field through $\mathcal{P}_4$ and $\mathcal{P}_5$, respectively. The term $\mathcal{P}_6$ may produce TME by extracting energy from the mean magnetic field. It may also produce MME by transferring energy from the turbulent magnetic field to the mean magnetic field.  Finally, the viscous dissipation components ($\mathfrak{D}$ and $\mathcal{D}$) convert KE to IE, while the ME is transformed to IE via Joule dissipation components ($\mathfrak{D}^M$ and $\mathcal{D}^M$). In the absence of a horizontally-averaged mean velocity field, as in the present problem, the energy pathways associated with MKE, marked in grey, are insignificant.

\begin{table*}
  \begin{center}
\def~{\hphantom{0}}
  \begin{tabular}{lccccccccccccccccc}
       $\text{case}$ & $\mathcal{R}$ & $Pm$& $Ro_C$  &  $\widetilde{Ra}$   &  $\widetilde{Rm}$ & $\langle\mathfrak{M}\rangle/M$ &  $M/E$ &  $Nu$ &  $\text{type}$ \\[3pt]
       R10Pm1N  & 10 & 1 & 0.069 & 76.0 & 33.4 & 0.0006 & 0.1749 & 55.0 & $\text{small}$\\
       R10Pm1F  & 10 & 1 & 0.074 & 87.0 & 25.9 & 0.0097 & 0.9350 & 64.7 & $\text{small}$\\       
       R10Pm0.1F & 10 & 0.1 & 0.074 & 87.0 & 3.2 & 0.2811 & 0.7605 & 60.7 & $\text{large}$\\
       R2Pm0.2N  & 2 & 0.2 & 0.031 & 15.2 & 1.5 & 0.6129 & 0.4866 & 8.5 & $\text{large}$\\       
       
  \end{tabular}
  \caption{Volume-averaged diagnostic quantities for the dynamo simulations at $E=5\times10^{-7}$ and $Pr=1$. The last column indicates the dynamo types.} 
  \label{tab:data}
   \end{center}
\end{table*}

\begin{figure}
\centering
(a)\includegraphics[width=0.46\linewidth,trim={0cm 0cm 0cm 0cm},clip]{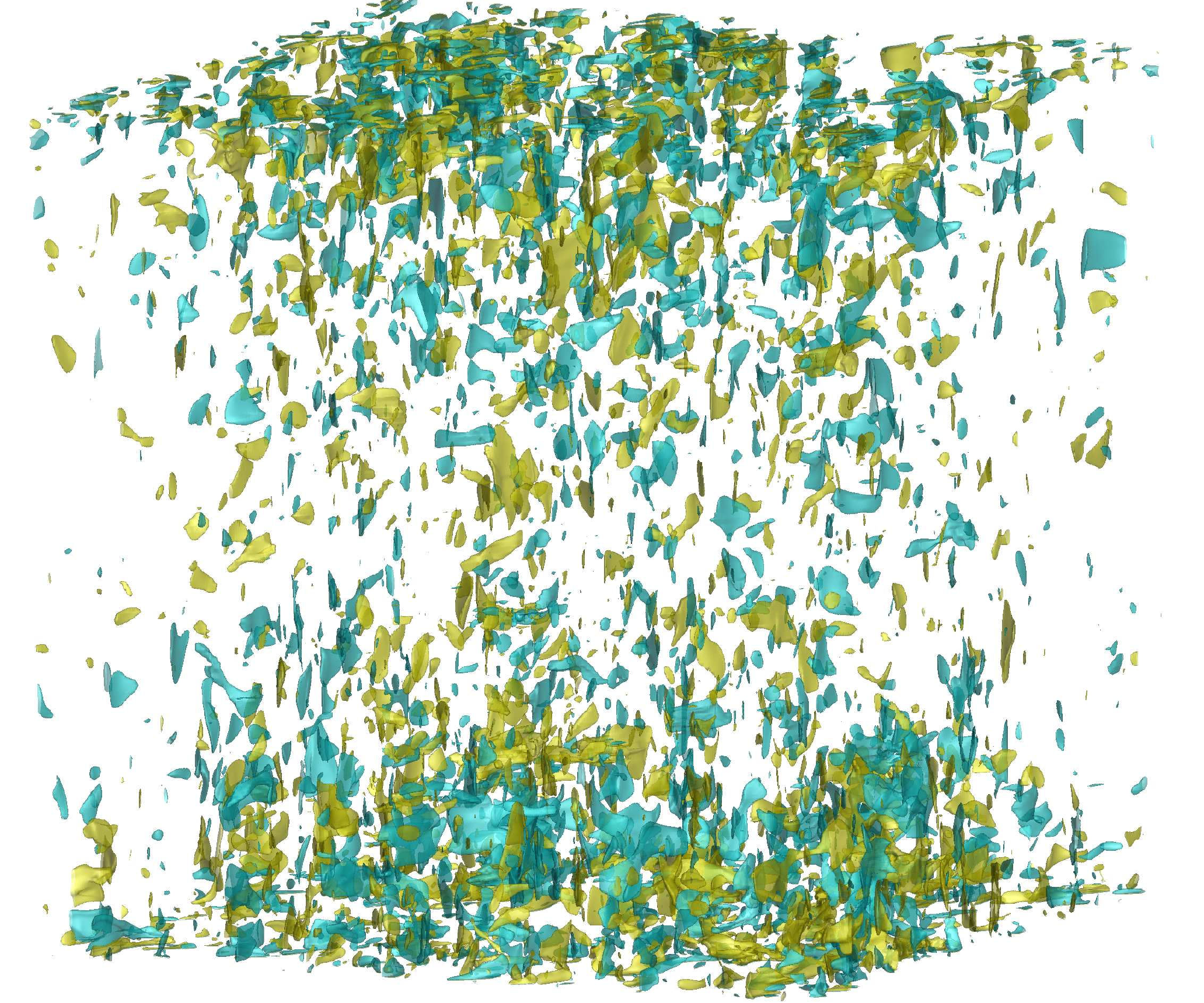}
(b)\includegraphics[width=0.46\linewidth,trim={0cm 0cm 0cm 0cm},clip]{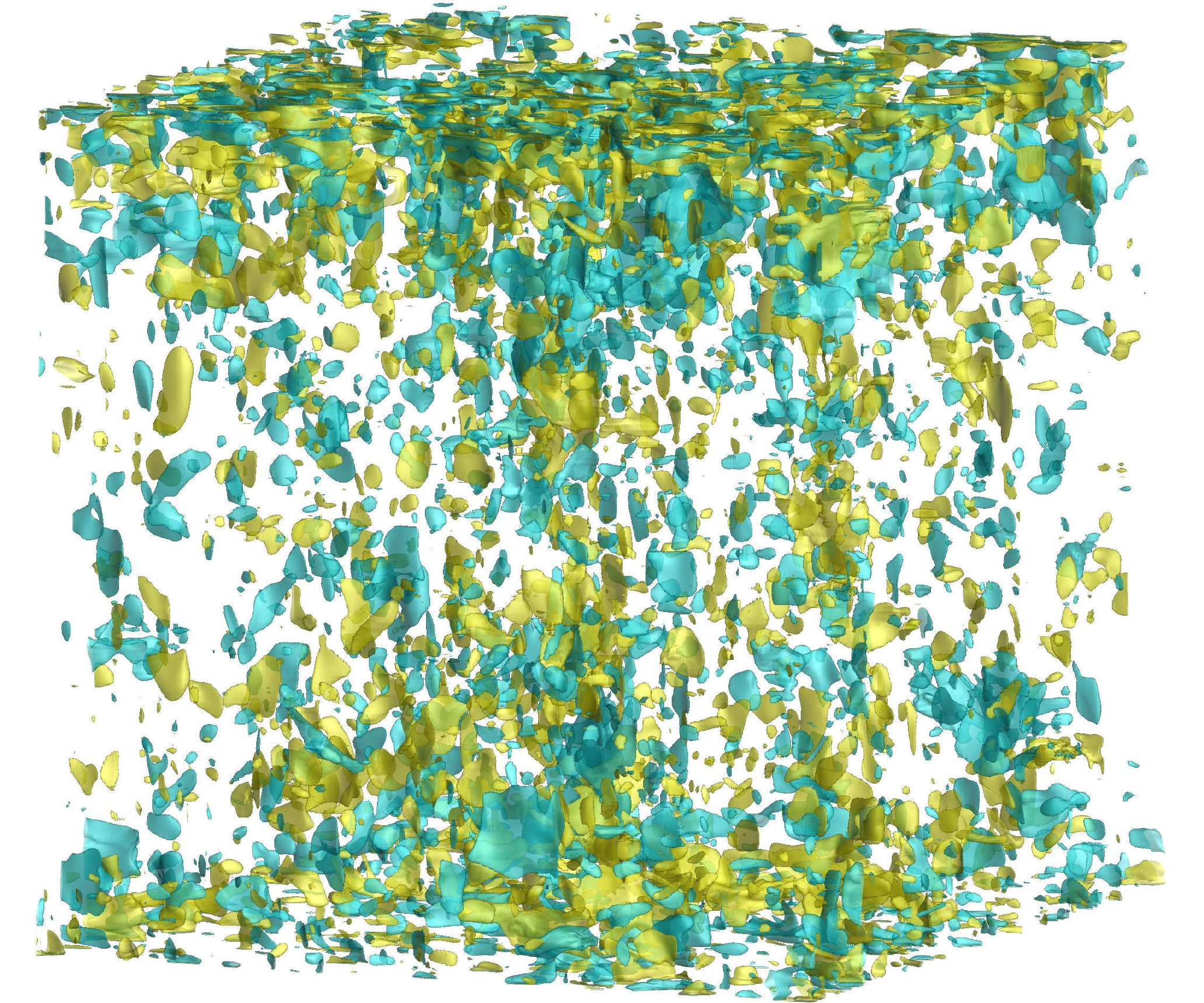}
(c)\includegraphics[width=0.46\linewidth,trim={0cm 0cm 0cm 0cm},clip]{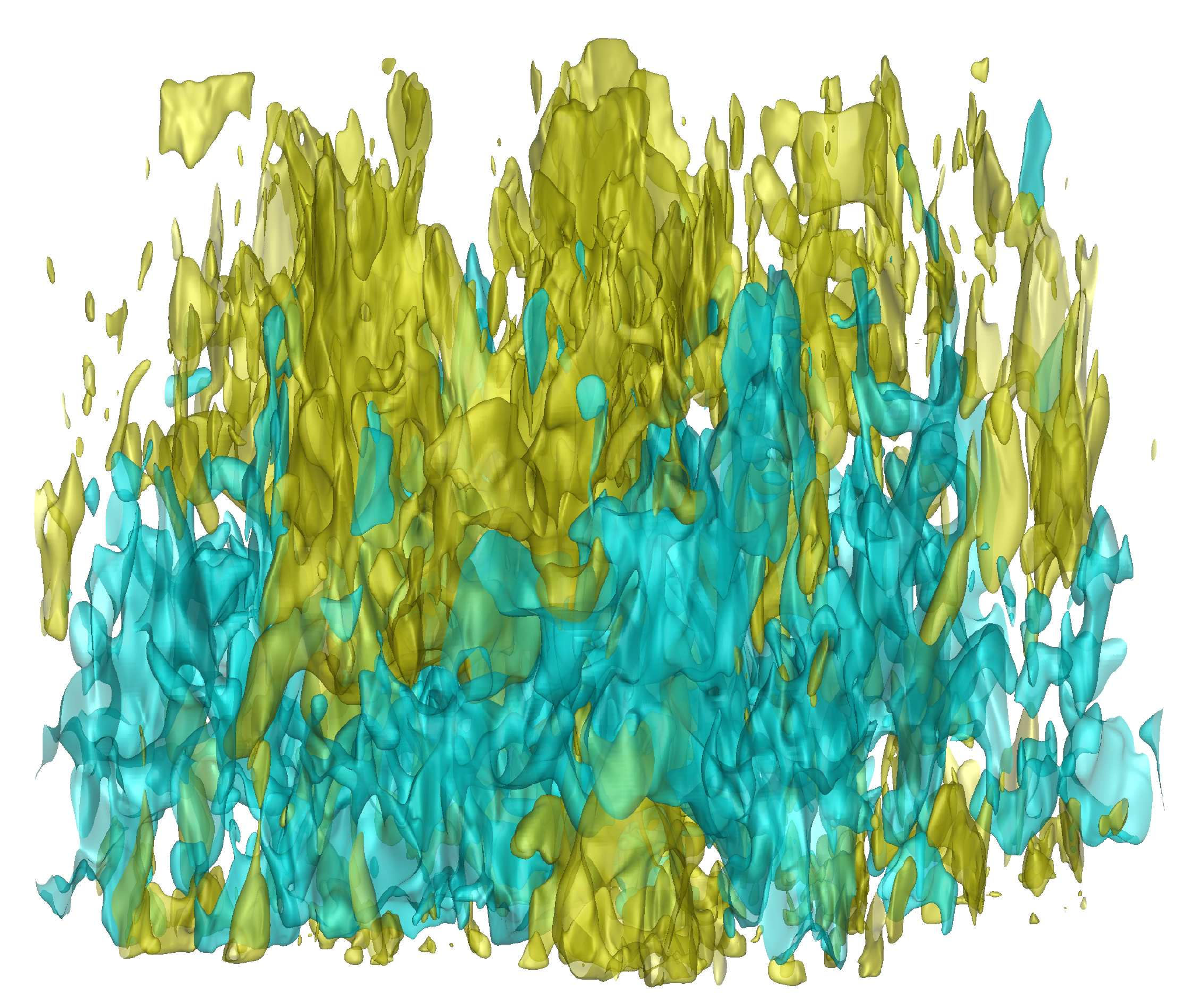}
(d)\includegraphics[width=0.46\linewidth,trim={0cm 0cm 0cm 0cm},clip]{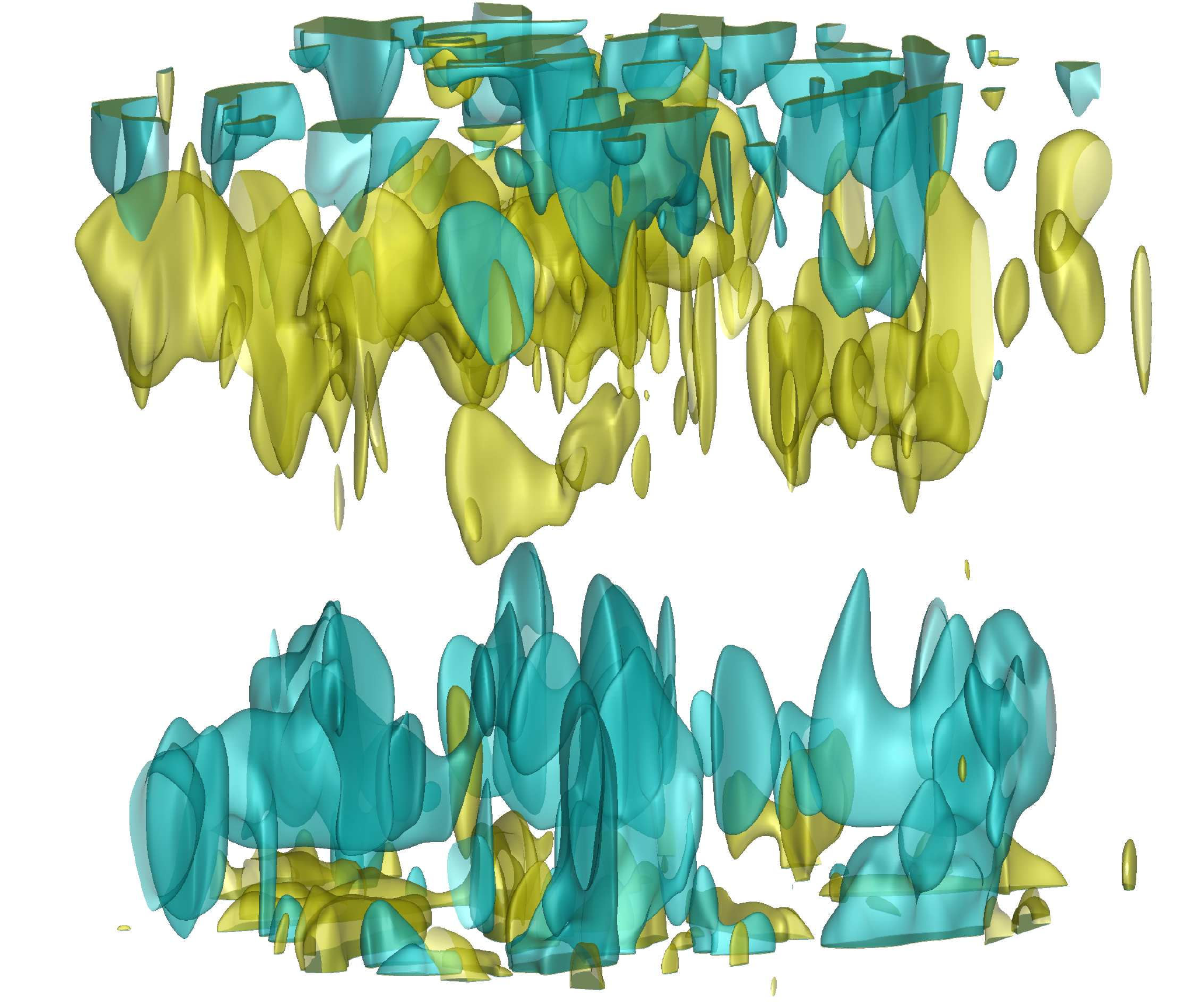}
\caption{The structure of the magnetic field generated by the dynamos for the cases (\textit{a}) R10Pm1F, (\textit{b}) R10Pm1N, (\textit{c}) R10Pm0.1F,(\textit{d}) R2Pm0.2N as visualized by the isosurface $B_{1}=\pm0.03$ (olive-positive, blue-negative)}
\label{fig:b1}
\end{figure}

\subsection{Numerical Details}\label{sec:numerical}
To characterize the different types of dynamos based on the energy pathway diagram in figure \ref{fig:energyflow}, we perform DNS of RBC-driven dynamos in a doubly-periodic domain of unit aspect ratio. We use $1024$ uniform grids in each horizontal direction ($x_1$ and $x_2$) and $256$ grids in the vertical direction ($x_3$) that are clustered near the boundaries. A finite-difference solver has been used for the simulations that have been validated extensively for studies on rotating stratified flow \citep{pal_2020}, and various transitional and turbulent shear flows \citep{Pal2013, Pal2015, pal_2020b}. Details of the solver, the validation studies, the rationale behind the choice of the grids, and the domain size are given in \citep{naskar_2022a,naskar_2022b}. We choose two dynamos at $\mathcal{R}=Ra/Ra_c=10$ and $Pm=1$ where $Ra_c$ is the Rayleigh number at the onset of non-magnetic rotating convection at $E=5\times10^{-7}$ \citep{naskar_2022b}. At this Ekman number, the critical Rayleigh number has the value $Ra_c=3.830\times10^9$ for no-slip and $Ra_c=2.192\times10^9$ for free-slip boundary conditions. In table \ref{tab:data}, the case R10Pm1N is simulated using no-slip and electrically conducting boundaries,  whereas free-slip and pseudo-vacuum boundaries \citep{naskar_2022b} are incorporated for the R10Pm1F case. The instantaneous snapshots of the magnetic field in the $x_1$-direction, as depicted in figures \ref{fig:b1}a and b, illustrate the small-scale nature of the magnetic field produced by these turbulent dynamos. Another turbulent dynamo case has been simulated, by lowering $Pm$ to $0.1$, following \citep{yan_2022b}, as denoted by R10Pm0.1F in table \ref{tab:data}. In this case, lowering $Pm$ leads to large-scale magnetic field generation, as evident from figure \ref{fig:b1}c. Additionally, we simulate a case with $\mathcal{R}=2$ and $Pm=0.2$ and designate it as R2Pm0.2N. This case is also a large-scale dynamo with weakly-nonlinear convection \citep{stellmach_2004} as demonstrated by the large-scale magnetic field in figure \ref{fig:b1}d. 

The volume-averaged diagnostic parameters reported in table \ref{tab:data} outline the global behaviour of these dynamos. The convective Rossby number, $Ro_C=E(Ra/Pr)^{1/2}$, representing the ratio of inertia to Coriolis force, is of the order $10^{-2}$, indicating the dominant role of the Coriolis force. Therefore, all the dynamos are produced by rapidly rotating convection with comparatively small inertia \citep{naskar_2022b}. The reduced Rayleigh number $\widetilde{Ra}=RaE^{4/3}$ indicates that the dynamos operate in a turbulent state of the flow for $\mathcal{R}=10$, whereas weakly-nonlinear columnar convection can be observed for $\mathcal{R}=2$ \citep{naskar_2022a,stellmach_2004}. The reduced magnetic Reynolds number $\widetilde{Rm}=RmE^{1/3}$ represents the strength of electromagnetic induction relative to ohmic diffusion at the convective scales. Large-scale dynamos are expected to be found for $\widetilde{Rm}\lesssim O(1)$ \citep{yan_2022b}. The distinction between large- and small-scale dynamos \citep{tilgner_2012} is apparent from the mean energy fraction $\langle\mathfrak{M}\rangle/M$, where $M=\langle\mathfrak{M}\rangle+\langle\mathcal{M}\rangle$. For the small-scale dynamos, the volume-averaged MME is three orders of magnitude smaller than the TME, while they are of the same order for the large-scale dynamos. Additionally, the fraction of magnetic energy $M/E$, where $E=M+K$ is the total energy, can be regarded as the efficiency of the dynamo action. 

\section{Results}\label{sec:results}
\begin{figure}
\centering
$\color{black}{\boldsymbol{\mathcal{B}}},$ \hspace{1mm} $\color{BrickRed}{\boldsymbol{\epsilon_v}},$ \hspace{1mm} $\color{ForestGreen}{\boldsymbol{\frac{\partial \mathcal{T}_j}{\partial x_j}}},$ \hspace{1mm} $\color{magenta}{\boldsymbol{\frac{\partial \mathcal{T}_p}{\partial x_j}}},$ \hspace{1mm} $\color{olive}{\boldsymbol{\frac{\partial \mathcal{T}_t}{\partial x_j}}},$ \hspace{1mm} $\color{Orchid}{\boldsymbol{\frac{\partial \mathcal{T}_v}{\partial x_j}}},$ \hspace{1mm} $\color{Cyan}{\boldsymbol{\frac{\partial \mathcal{T}_M}{\partial x_j}}},$ \hspace{1mm} $\color{blue}{\boldsymbol{\mathcal{P}}},$ \hspace{1mm} $\color{blue}{\boldsymbol{-\mathcal{P}_1 ( \circ )},}$ \hspace{1mm} $\color{blue}{\boldsymbol{-\mathcal{P}_2 ( \square )},}$ \hspace{1mm} $\color{blue}{\boldsymbol{-\mathcal{P}_3 ( \diamond )}}$
(a)\includegraphics[width=0.46\linewidth,trim={0cm 1cm 0cm 0cm},clip]{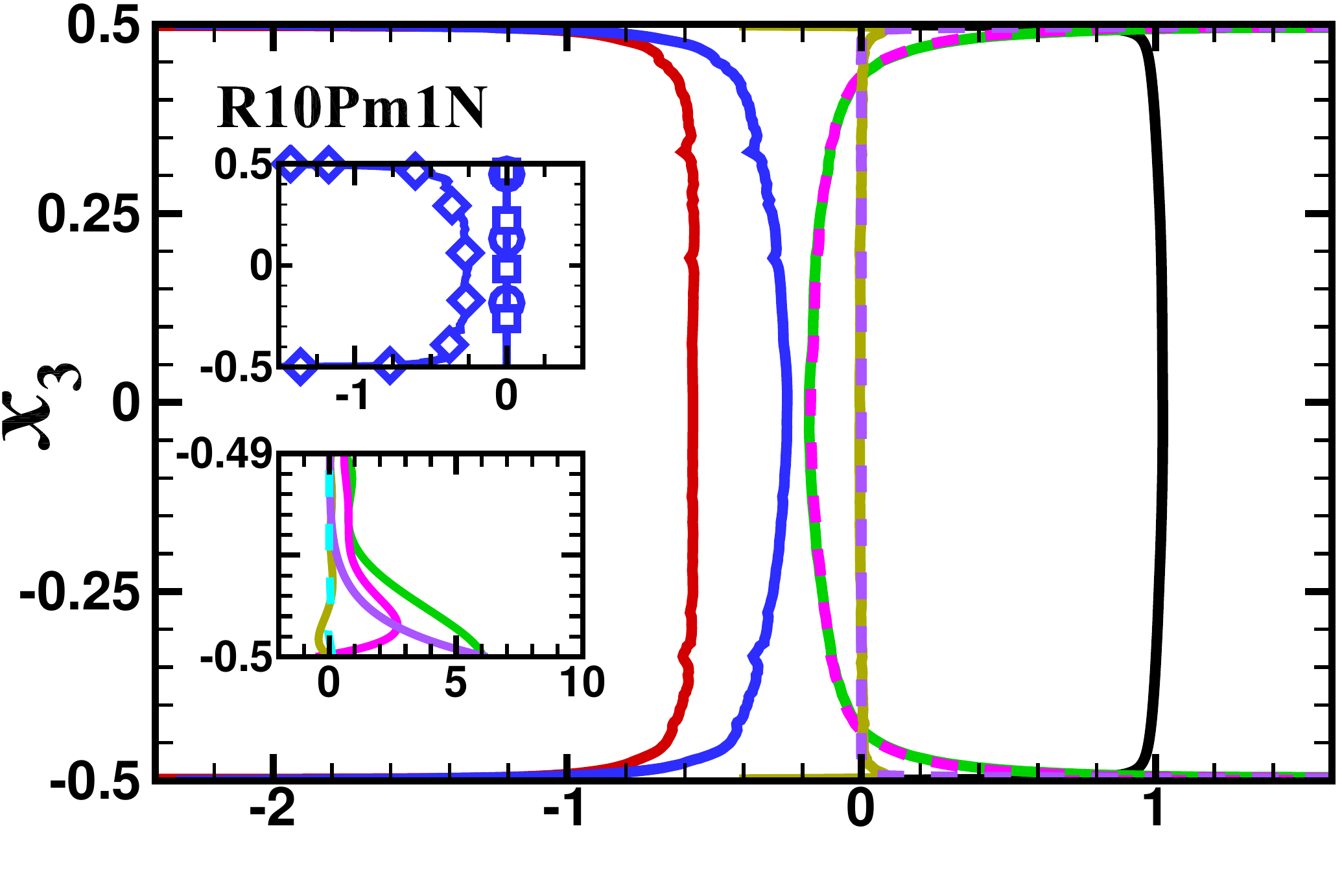}
(b)\includegraphics[width=0.46\linewidth,trim={0cm 1cm 0cm 0cm},clip]{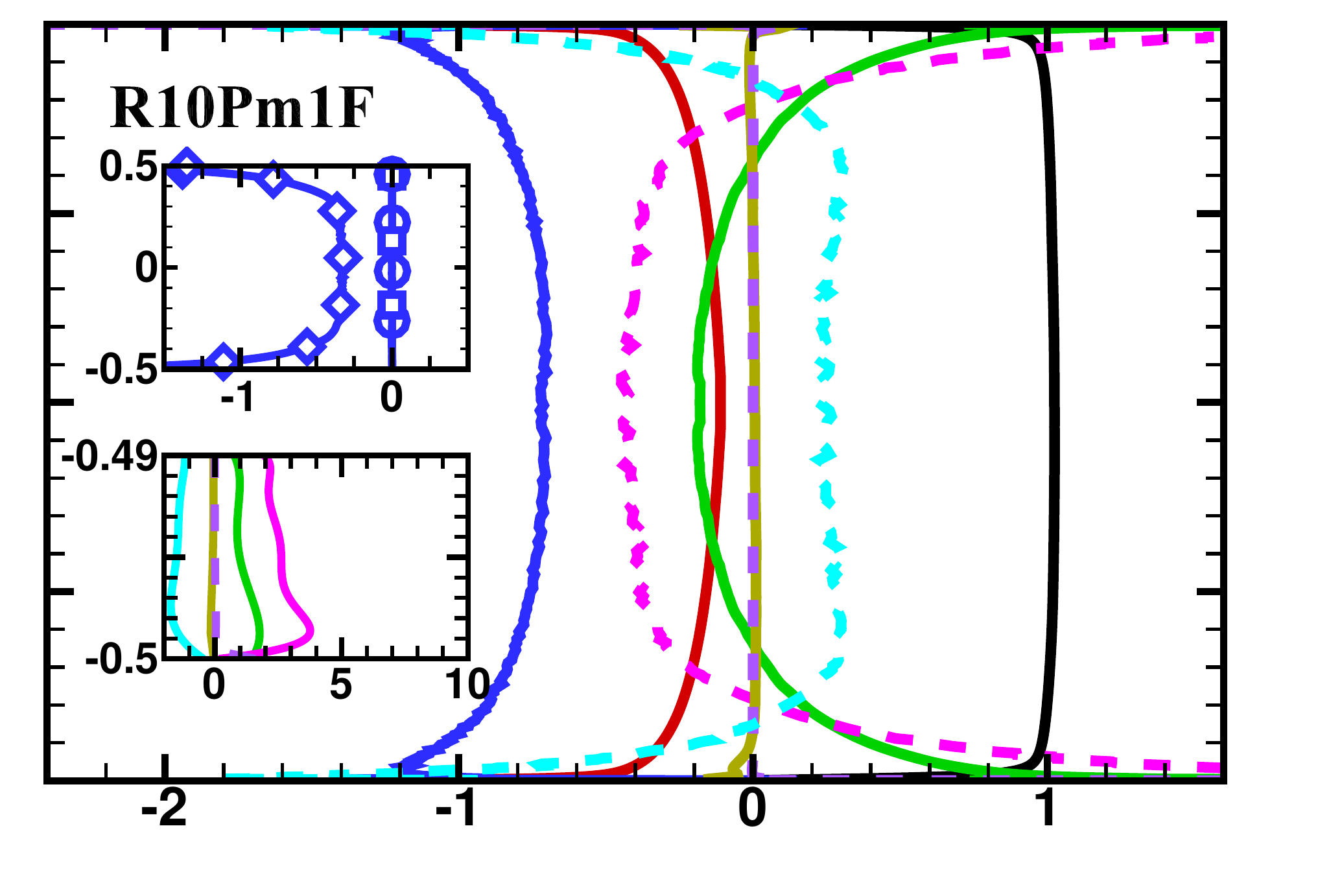}\\
(c)\includegraphics[width=0.46\linewidth,trim={0cm 0cm 0cm 0cm},clip]{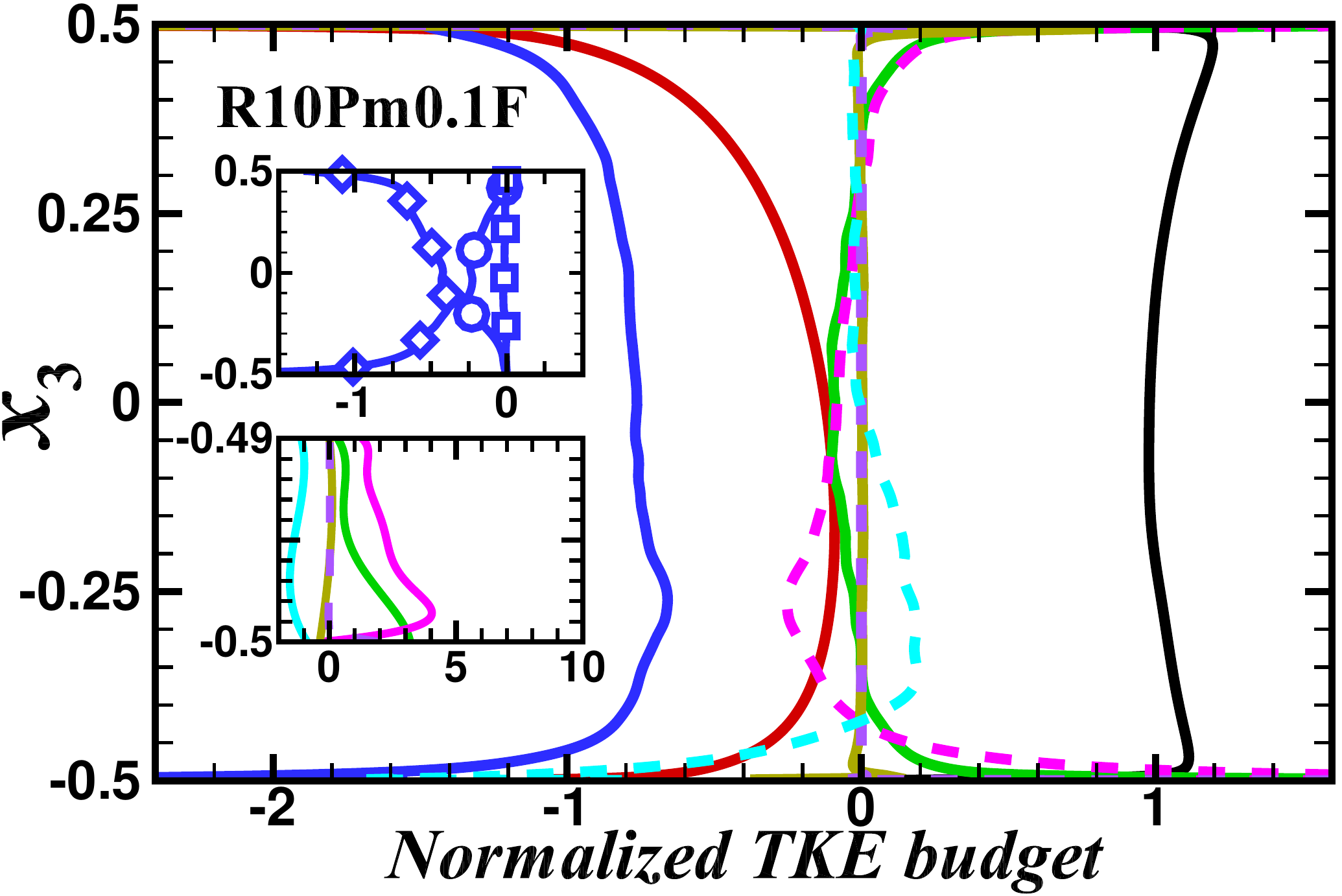}
(d)\includegraphics[width=0.46\linewidth,trim={0cm 0cm 0cm 0cm},clip]{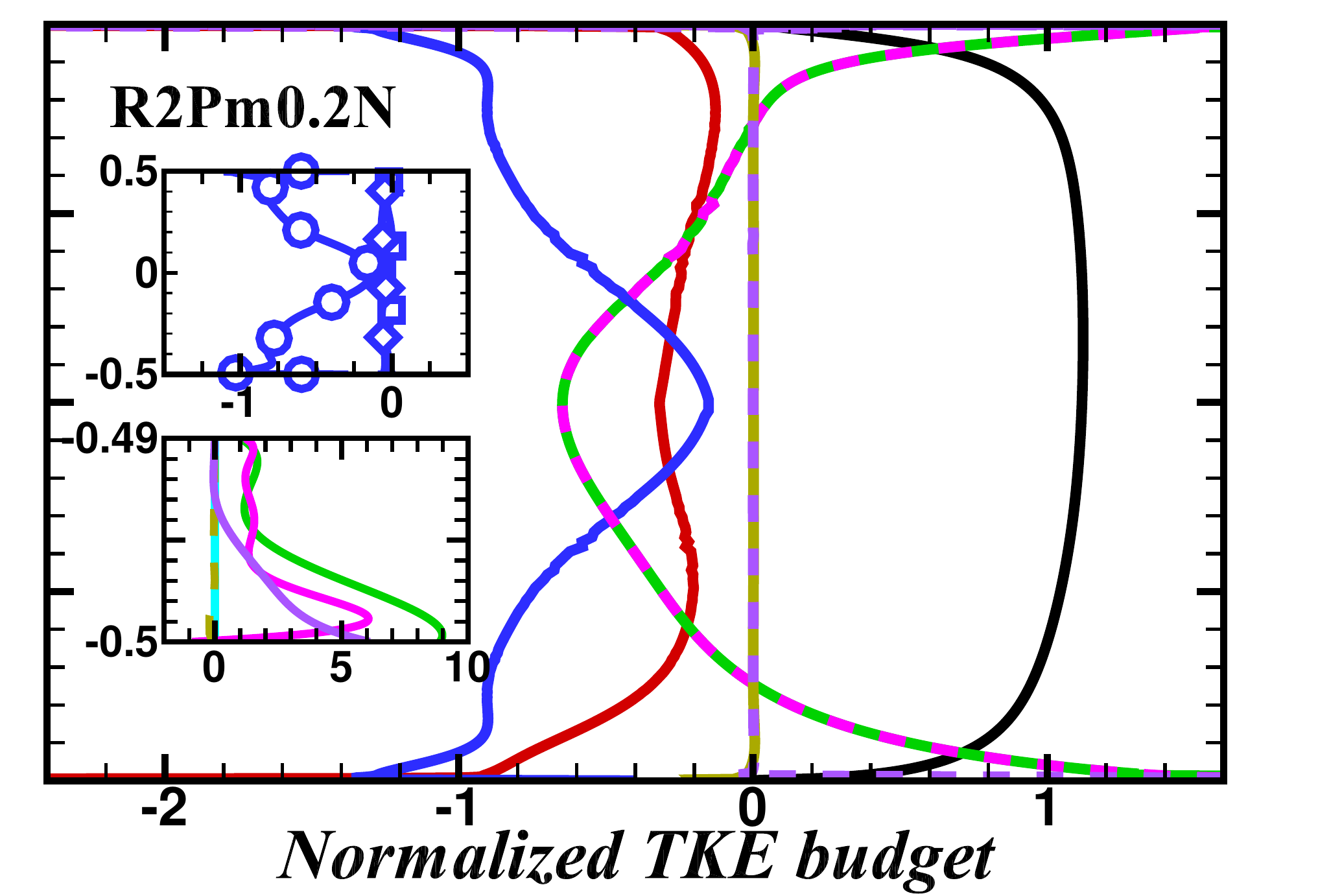}
\caption{Vertical variation of the terms in TKE budget  for (\textit{a}) R10Pm1F, (\textit{b}) R10Pm1N, (\textit{c}) R10Pm0.1F ,(\textit{d}) R2Pm0.2N , cases. The insets in the top left corners show the components of the energy exchange $\mathcal{P}$. The insets in the bottom left corners shows a breakup of the transport terms $\partial\mathcal{T}_{j}/\partial x_j$}
\label{fig:tke}
\end{figure}

Figure \ref{fig:tke} demonstrates the vertical variation of the horizontally averaged TKE terms in equation \ref{eqn:TKE_budget} for the four cases. All the terms in this equation have been averaged over time, and normalized by $(RaPr)^{1/2}/(Nu-1)$ in the figure, so that the volume average of the source term ($\mathcal{B}$) is unity\citep{Kerr_2001}. The Nusselt number $Nu=qd/k\Delta T$, where q is the total vertical heat flux, is a non-dimensional measure of convective heat transport through the fluid layer, as reported in table \ref{tab:data}. At a statistically stationary state, there is a primary balance among the turbulent buoyancy flux ($\mathcal{B}$), TKE transport ($\partial \mathcal{T}_j/\partial x_j$),  viscous dissipation ($\mathcal{D}$) and the conversion to magnetic energy ($\mathcal{P}$) for all the dynamos. The individual components of $\mathcal{P}$ and $\partial \mathcal{T}_j/\partial x_j$ are shown in the top and the bottom insets, respectively. For instance, in figure \ref{fig:tke}a, the TKE is generated by $\mathcal{B}$ in the bulk, which is partly transported by $\partial \mathcal{T}_j/\partial x_j$ towards the boundaries where $\mathcal{D}$ dominates. Among the transport components, the pressure transport ($\partial \mathcal{T}_P/\partial x_j$) is the primary mechanism that transfers TKE towards boundaries while the viscous transport is higher near the boundaries as observed in the bottom inset, similar to non-rotating non-magnetic RBC \citep{petschel_2015}. The vertical variation of all the terms, except $\mathcal{P}$, are qualitatively similar with non-magnetic rotating convection \citep{kunnen_2009,guzman_2020}. 

\begin{table*}
  \begin{center}
\def~{\hphantom{0}}
  \begin{tabular}{lcccccccccc}
       $\text{case}$ & $\langle\mathcal{P}_1\rangle$ &  $-\langle\mathcal{P}_2\rangle$ & $\langle\mathcal{P}_3\rangle$ &  $-\langle\mathcal{P}_6\rangle$ & $\langle\mathcal{D}\rangle$ & $\langle\mathcal{D}^{M}\rangle$ & $\langle\mathfrak{D}^{M}\rangle$ &  $\text{type}$ \\[3pt]
       R10Pm1N  & 0.0002 & 0.0000 & 0.3491 & 0.0000 & 0.6232 & 0.3507 & 0.0000 & $\text{small}$\\
       R10Pm1F  & 0.0004 & 0.0014 & 0.7549 & 0.0013 & 0.1627 & 0.7612 & 0.0001 & $\text{small}$\\       
       R10Pm0.1F & 0.1397 & 0.0093 & 0.5696 & 0.0069 & 0.2457 & 0.7189 & 0.0026 & $\text{large}$\\
       R2Pm0.2N  & 0.3399 & 0.0068 & 0.0538 & -0.0055 & 0.6252 & 0.3878 & 0.0120 & $\text{large}$\\       
       
  \end{tabular}
  \caption{Volume-averaged diagnostic quantities for the dynamo simulations at $E=5\times10^{-7}$ and $Pr=1$. The last column indicates the dynamo types.} 
  \label{tab:budget}
   \end{center}
\end{table*}

The small-scale dynamos R10Pm1N and R10Pm1F in figure \ref{fig:tke}a and \ref{fig:tke}b differ only in the boundary conditions. The choice of a no-slip boundary condition in R10Pm1N results in an Ekman layer near the boundaries. Moreover, the perfectly conducting magnetic boundary condition constrains the magnetic field to be horizontal near the boundaries as compared to a vertical magnetic field at the boundaries for a pseudo-vacuum boundary condition in R10Pm1F \citep{naskar_2022b}. For both of these dynamos, the small-scale production of magnetic energy ($\mathcal{P}_3$) is the dominant component of $\mathcal{P}$. For R10Pm1N, the amount of TKE converted to IE via $\mathcal{D}$ is higher than the production of magnetic energy ($\mathcal{P}$), whereas the latter is higher for R10Pm1F (see also table \ref{tab:budget}). Another difference between these dynamos arises from the transport of TKE due to magnetic energy, which is dominated by the small-scale flux $\overline{u^{\prime}_iB^{\prime}_iB^{\prime}_j}$, while the large-scale flux $\overline{B_j}\overline{u^{\prime}_i{}B^{\prime}_i}$ remains small. The small-scale magnetic field transports TKE towards the interior from the boundaries in R10Pm1F, while the same term is small for R10Pm1N. The viscous transport dominates the other transport terms near the boundaries for R10Pm1N. However, this term is small near the boundaries in R10Pm1F owing to the absence of an Ekman layer.

A comparison between the small-scale dynamo R10Pm1F with the large-scale dynamo R10Pm0.1F reveals significant differences in the energy pathways. The mean magnetic field plays an important role in converting TKE to TME through the term $\mathcal{P}_1$ in R10Pm0.1F, while this term is insignificant for R10Pm1F. However, for both cases, the redistribution of TKE by the magnetic field plays a part in the budget with a major contribution from the small-scale flux  $\overline{u^{\prime}_iB^{\prime}_iB^{\prime}_j}$ in R10Pm1F, while the large-scale flux $\overline{B_j}\overline{u^{\prime}_i{}B^{\prime}_i}$ is significant for R10Pm0.1F. Additionally, we can contrast the large-scale turbulent dynamo R10Pm0.1F against a large-scale dynamo with weakly non-linear convection R2Pm0.2N in figure \ref{fig:tke}. For R2Pm0.2N, the conversion of TKE to TME occurs through $\mathcal{P}_1$ while the small-scale production of TME, $\mathcal{P}_3$ remains small. Among these four dynamos, the part of TKE that converts to IE due to $\mathcal{D}$ is higher, as compared with the conversion to magnetic energy via $\mathcal{P}$, when no-slip boundary condition is used (see table \ref{tab:budget}). The conversion to magnetic energy $\mathcal{P}$ is higher with free-slip conditions, which makes R10Pm1F and R10Pm0.1F more efficient dynamos with higher magnetic energy fraction $M/E$, as compared to the no-slip boundary condition. Also, the transport of TKE by the magnetic field is significant only for pseudo-vacuum magnetic boundary conditions.

\begin{figure}
\centering
$\textbf{solid lines:}$ \hspace{2mm}
$\color{ForestGreen}{\boldsymbol{\mathcal{P}_1}},$ \hspace{1mm} $\color{black}{\boldsymbol{\mathcal{P}_3}},$ \hspace{1mm} $\color{Orange}{\boldsymbol{\mathcal{D}^{M}}},$ \hspace{1mm} $\color{Orchid}{\boldsymbol{-\frac{\partial \mathcal{T}^{M}}{\partial x_j}}}$ \hspace{1mm} $\textbf{dashed lines:}$ \hspace{2mm} $\color{blue}{\boldsymbol{-\mathcal{P}_2}},$ \hspace{1mm} $\color{magenta}{\boldsymbol{\mathcal{P}_6}},$ \hspace{1mm} $\color{BrickRed}{\boldsymbol{\mathfrak{D}^{M}}},$ \hspace{1mm} $\color{olive}{\boldsymbol{-\frac{\partial \mathfrak{T}^{M}}{\partial x_j}}}$
(a)\includegraphics[width=0.46\linewidth,trim={0cm 0cm 0cm 0.5cm},clip]{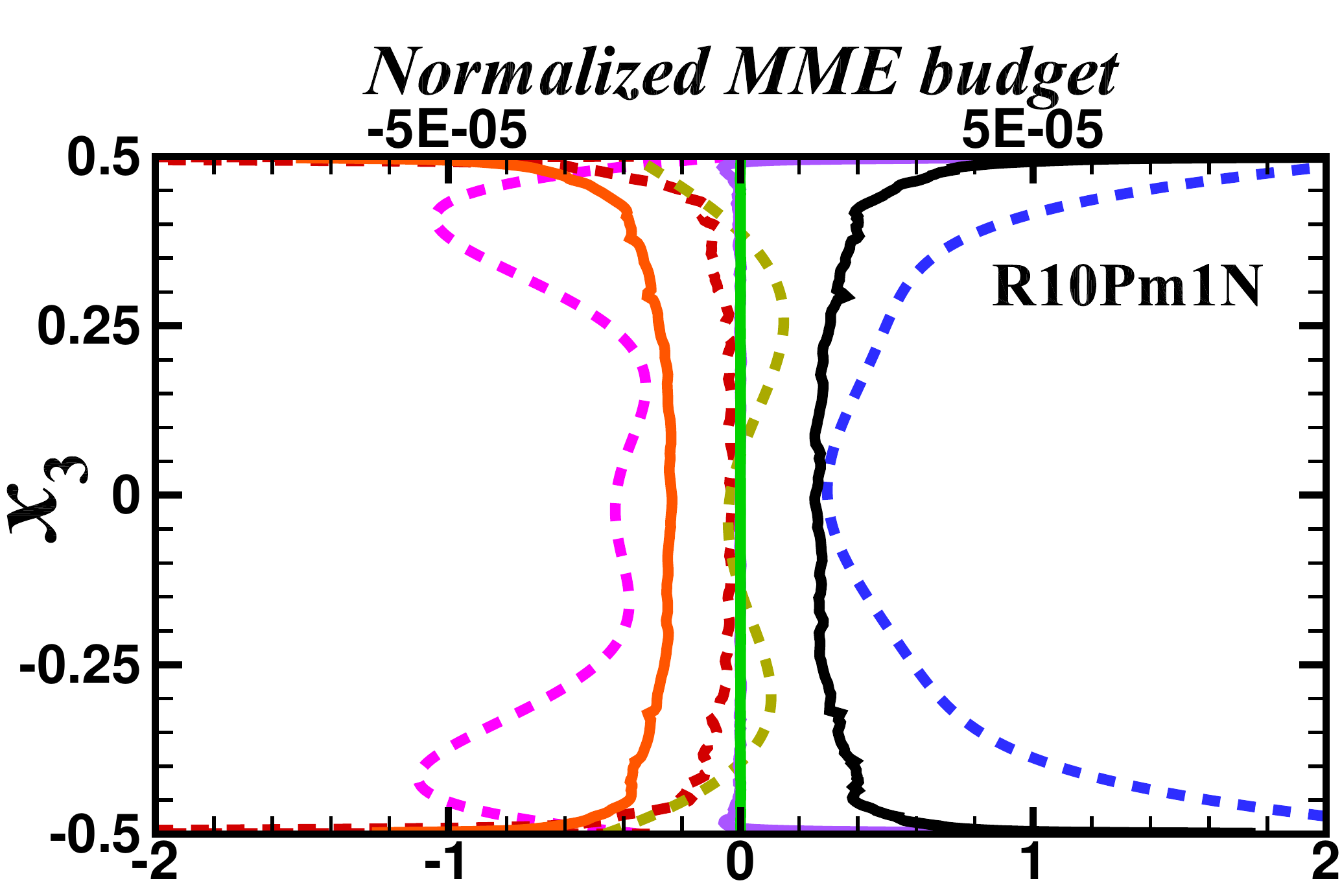}
(b)\includegraphics[width=0.46\linewidth,trim={0cm 0cm 0cm 0.5cm},clip]{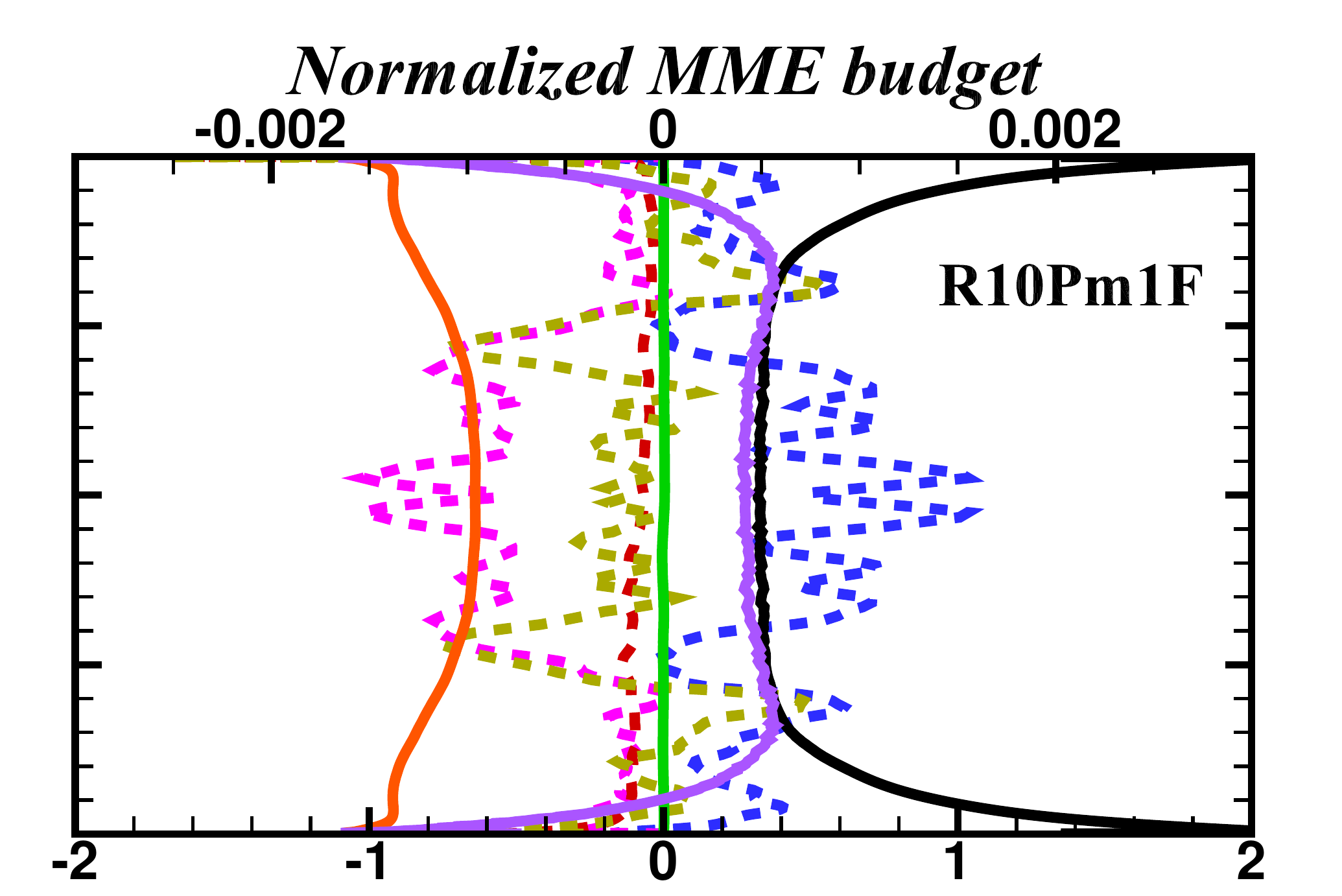}\\
(c)\includegraphics[width=0.46\linewidth,trim={0cm 0cm 0cm 0.5cm},clip]{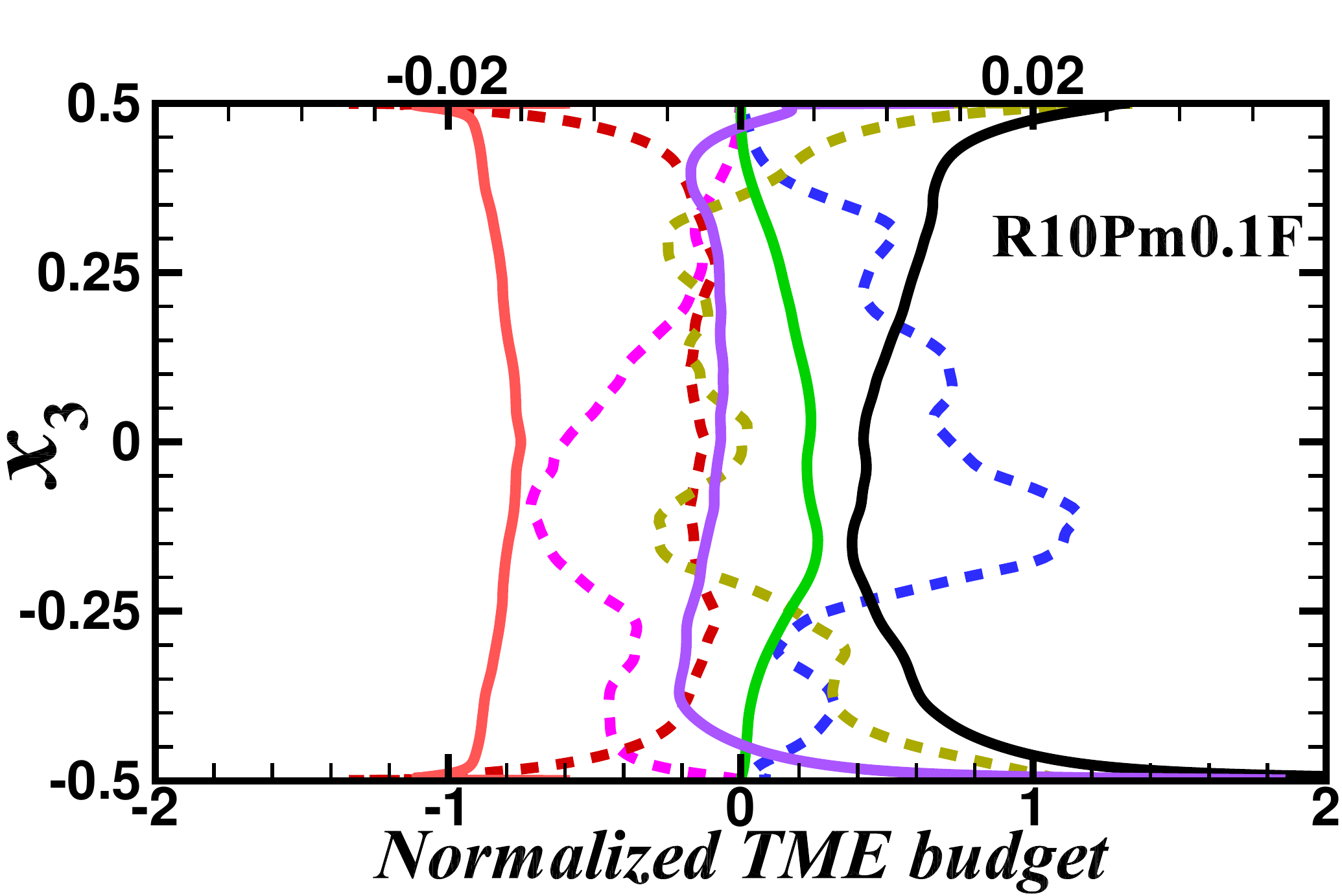}
(d)\includegraphics[width=0.46\linewidth,trim={0cm 0cm 0cm 0.5cm},clip]{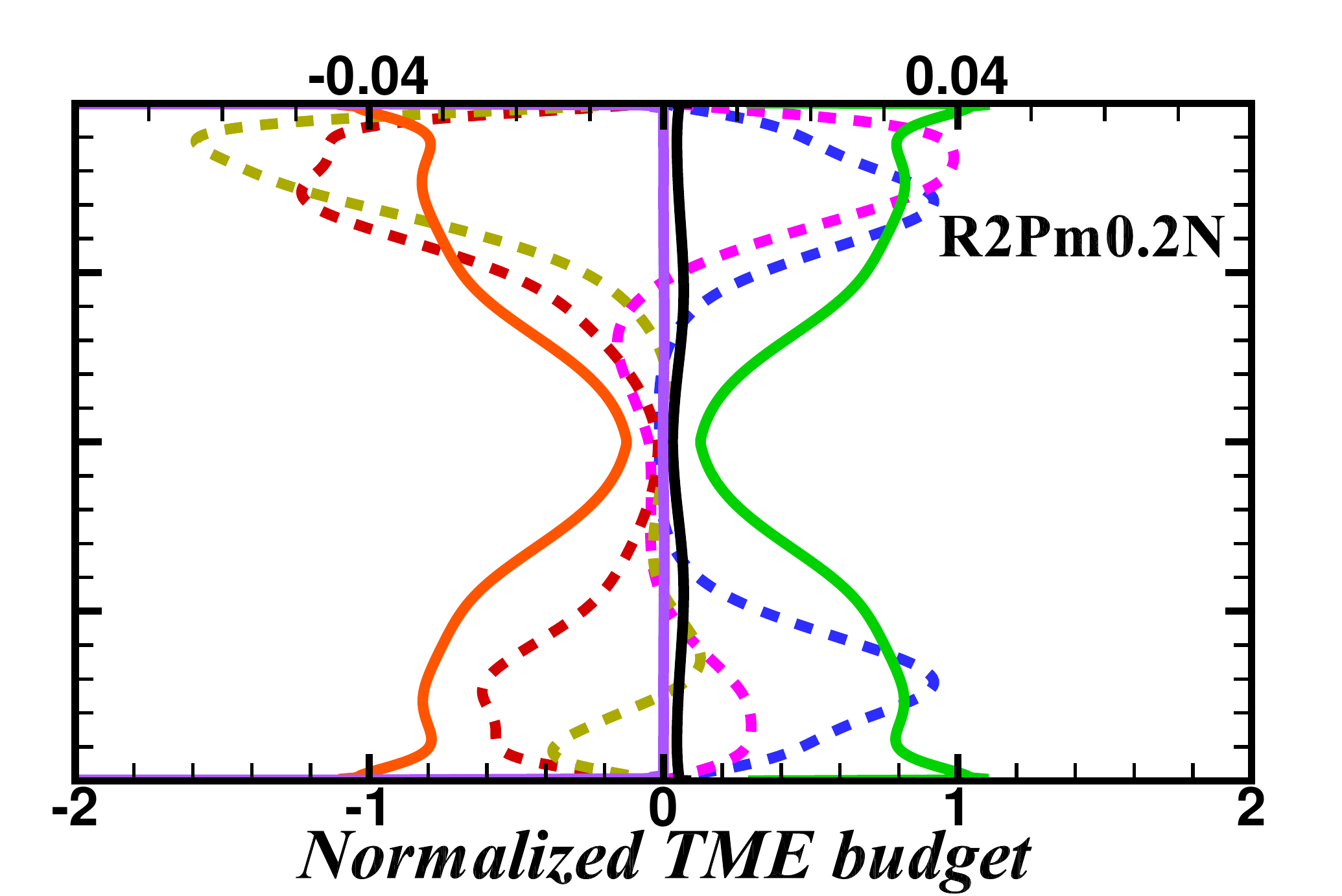}
\caption{Vertical variation of the terms in TME and MME budgets  for (\textit{a}) R10Pm1F, (\textit{b}) R10Pm1N, (\textit{c}) R10Pm0.1F,(\textit{d}) R2Pm0.2N cases. TME(MME) budget terms are plotted using solid(dotted) lines and are to be read from the bottom(top) abscissa.}
\label{fig:tme}
\end{figure}

The magnetic energy balance of the dynamos is presented in figure \ref{fig:tme}. The solid lines represent the terms in the TME budget and are plotted at the bottom abscissa. The terms of the MME budget are represented by the dashed lines and are plotted at the top abscissa.  The energy flow direction of each dynamo in figure \ref{fig:energyflow} becomes apparent with the help of table \ref{tab:budget} and figure \ref{fig:tme}. The terms $\mathcal{P}_1$ and $\mathcal{P}_3$ are positive, representing a conversion of TKE to TME by the action of a large-scale and a small-scale magnetic field, respectively. The term $\mathcal{P}_2$ is negative, indicating a generation of MME at the expense of TKE. The term $\mathcal{P}_6$ has negative values indicating a conversion of MME to generate TME, except for the case R2Pm0.2N where the turbulent magnetic field provides energy to the mean magnetic field, exhibiting an upscale transfer of energy.

In figure \ref{fig:tme}a the small-scale dynamo R10Pm1N has a primary balance between the small-scale production $\mathcal{P}_3$ and the Joule dissipation $\mathcal{D}^{M}$. In comparison to these TME terms, the MME terms are three orders of magnitude smaller. The part of TKE that converts to MME ($\mathcal{P}_2$) again transforms to TME, with a small mean dissipation $\mathfrak{D}^{M}$. A similar energy conversion is observed in R10Pm1F in figure \ref{fig:tme}b, though the MME budget terms are one order of magnitude higher than R10Pm1N. Unlike R10Pm1N, the transport of TME $\partial \mathcal{T}^{M}_j/\partial x_j$ plays a significant role in the balance in R10Pm1F by redistributing the TME from the boundaries towards the interior of the domain. For the large-scale dynamos R10Pm0.1 in figure \ref{fig:tme}c, the MME terms increase by another order of magnitude compared to the small-scale dynamo R10Pm1F. Further, the large-scale production of TME term $\mathcal{P}_1$ now makes a significant contribution to the budget in the interior of the domain. In the MME budget, $\mathcal{P}_2$ is partially balanced by $\mathcal{P}_6$ in the interior. However, the rest of the MME is transported towards the boundary and converted to IE by Joule dissipation $\mathfrak{D}^M$. The large-scale dynamo in figure \ref{fig:tme}d demonstrates a transfer of TME to MME by $\mathcal{P}_6$ near the boundaries. This is the only dynamo where both the small-scale velocity and magnetic fields provide energy to the mean magnetic field through $\mathcal{P}_2$ and $\mathcal{P}_6$. Additionally, the large-scale production $\mathcal{P}_1$ is the primary source of TME generation whereas $\mathcal{P}_3$ remains small in this case, in contrast to R10Pm0.1F in figure \ref{fig:tme}c.

\section{Conclusions}\label{sec:conclusion}

In summary, we have performed direct numerical simulations of four dynamos to compare their magnetic and kinetic energy budgets. The small-scale dynamos R10Pm1N and R10Pm1F differ by the relative magnitude of small-scale magnetic energy production $\mathcal{P}_3$ and viscous dissipation $\mathcal{D}$, with the latter being higher for R10Pm1N. This indicates comparatively efficient dynamo action with free-slip, pseudo-vacuum boundaries that also promote the redistribution of TKE by the magnetic field, unlike a dynamo with no-slip, perfectly conducting boundaries. Nevertheless, the mean Joule dissipation is small for small-scale dynamos. The mechanism for transforming TKE to TME differs between a large- and a small-scale turbulent dynamo, with the large-scale production of TME $\mathcal{P}_1$ playing a significant role in the former. This large-scale production $\mathcal{P}_1$ becomes the dominant mechanism of TME production in the weakly non-linear dynamo R2Pm0.2N. MME is produced from TKE via the term $\mathcal{P}_2$ in the presence of a mean magnetic field gradient. For R2Pm0.2N, an upscale transfer of energy occurs through $\mathcal{P}_6$, which produces MME at the expense of TME. The scaling of these energy budget terms in the limit of small viscous and inertial forces, following \citep{calkins_2015}, should provide valuable insights into the mechanism of energy conversion in astrophysical dynamos. Additionally, a shell-to-shell energy transfer analysis \citep{guzman_2020} may elucidate further details on the mechanism of large-scale magnetic field generation. 




\backsection[Declaration of interests]{ The authors report no conflict of interest.}



\backsection[Author contributions]{  The authors contributed equally to analyzing data and reaching conclusions, and in writing the paper.}

\bibliographystyle{jfm}
\bibliography{jfm}

\end{document}